\documentclass[dvipdfmx,a4j,11.5pt]{article}
\usepackage[dvipdfmx]{graphicx}
\usepackage{color}
\usepackage{ascmac}
\usepackage{amsmath}
\usepackage{amsfonts}
\usepackage{braket}
\usepackage{url}
\usepackage{typearea}
\typearea{12}

\usepackage{graphicx}
\usepackage{braket}

\usepackage{color}

\begin{document}


\title{Bifurcation structure of unstable periodic orbits
  in plane Couette flow with the Smagorinsky model}

\author{Eiichi Sasaki\thanks{
 Graduate School of Engineering Science, 
 Akita University, 
 Akita, Japan 
 esasaki@gipc.akita-u.ac.jp}, Genta Kawahara\thanks{
Graduate School of Engineering Science, 
Osaka University, 
Osaka, Japan %
} 
 and Javier Jim\'enez\thanks{
School of Aeronautics, U. Polit\'ecnica de Madrid, 28040 Madrid, Spain
}}%

\date{\today}

\maketitle

\begin{abstract}
 In order to obtain insights into dynamics of developed plane Couette turbulence,
 this paper considers bifurcation structure of unstable periodic orbits (UPOs)
 in the large-eddy-simulation (LES) system 
 with the Smagorinsky-type eddy viscosity model.
 Treating the Smagorinsky constant as a bifurcation parameter,
 we detect the bifurcation points connecting two known UPOs 
 which were separately discovered in the Navier--Stokes system 
 [G. Kawahara \& S. Kida, J. Fluid Mech., {\bf 449}, 291--300 (2001)].
 At the moderately high Reynolds number,
 the LES UPO of the present study possesses the spanwise vortices which
 seem to be caused by the streak instability and appear in the central region of the channel.
 We note that  to our knowledge a spanwise vortex has not been reported through the UPO analysis of wall flows.
 The stretched spanwise vortices of the LES UPO enhance transfer 
 of the streamwise turbulent momentum as in developed near-wall turbulence.
\end{abstract}

\section{Introduction}

Understanding developed Navier--Stokes (NS) turbulence is one of the key issues in the dynamical systems theory. 
The theory tells us that 
a countably infinite number of unstable periodic orbits (UPOs), to be exact invariant sets, 
are embedded in a chaotic attractor \cite{2013_Hirsch}.
The stable and unstable manifolds of UPOs play roles of attractor and repeller for chaotic orbits,
so the chaotic orbits wander around these UPOs like a pin ball in phase space. 
Therefore, UPOs represent dynamical properties of  the chaotic attractor \cite{2003_Kato,2005_Kawasaki}.
For the NS system, the solution orbits of a turbulent state pass through the neighborhood of UPOs, and
the dynamical and structural properties of UPOs
could describe representative motions and vortex structures of turbulence. 
Using the UPO analysis, we could obtain theoretical descriptions of the important events in developed turbulence.
Here, we try to unveil essential dynamics of developed plane Couette turbulence through UPOs.

The approach based on the dynamical systems theory has succeeded especially in the characterization of
 low-Reynolds-number turbulence.
The turbulent motions are sustained by creation and subsequent breakdown of the streaks; 
the so-called regeneration cycle \cite{1991_Jimenez,1995_Hamilton}.
In the early days of UPO analysis, a steady state was discovered \cite{1990_Nagata,1997_Clever,1998_Waleffe}, 
and this solution is sometime called NCBW from the authors' names. 
We add that steady states are special cases of UPO without a period.
The flow structure of NCBW consists of a wavy low-velocity streak and a pair of streamwise vortices, 
resembling  the buffer-layer coherent structures in the near-wall region.
Kawahara \& Kida \cite{2001_Kawahara} examined plane Couette flow and found out two UPOs, one of which describes the regeneration cycle.
Viswanath \cite{2007_Viswanath} presented other UPOs also explaining the regeneration cycle.
In the past, a large number of UPOs reproducing the buffer-layer coherent structures have been reported
and the review is given by Kawahara et al. \cite{2012_Kawahara}.
For transition to turbulence in plane Couette flow of a large domain, 
turbulent spots and stripes are observed \cite{1995_Dauchot,2010_Duguet}.
Their flow structures consist of spatially localized patches of unsteady complex flow within a background of laminar flow and they eventually decay.
Brand \& Gibson \cite{2014_Brand} reported an unstable steady state, 
which characterizes the localized coherent structure, and its stability depicts the transient growth and decay of turbulence.
In addition, spatially localized solutions for wall-bounded shear flow have been found in pipe flow \cite{2013_Avila}, 
channel flow \cite{2014_Zammert}, and square-duct flow \cite{2015_Okino}.
Lustro et al. \cite{2019_Lustro} identified the first appearance of a homoclinic tangency in plane Couette flow, 
which means the onset of transiently turbulent state that finally laminerizes.
Using the rescaling of the Reynolds number, Eckhardt \& Zammert \cite{2018_Eckhardt} reported steady states and traveling waves, 
which localize in the wall-normal direction as the Reynolds number increases.
Concerning turbulence control using dynamical systems theory 
Kawahara \cite{2005_Kawahara} demonstrated laminarization strategy for plane Couette flow. 
Recently, Linkmann et al. \cite{2020_Linkmann} applied this control protocol for channel flow.
Note that we revisit work by Kawahara \& Kida \cite{2001_Kawahara} later.

Our interest is in developed turbulence.
When we use the conventional methods for the UPO analysis, 
we need ``good'' initial guesses which approximate the orbits of the target UPOs.
In practice, the good initial guesses have been obtained through heuristic methods. 
However, it is hard to identify turbulent orbits approaching UPOs at high Reynolds numbers.
This study considers UPOs found at low Reynolds numbers, and 
we track their branches by increasing the Reynolds number.

The high degrees of freedom of developed turbulence are another difficulty for the UPO analysis.
In order to decrease the degrees of freedom
the present study employs the eddy viscosity model in the large-eddy simulation (LES). 
In the LES system, the large scales are computed on the resolved grid points, while the small-scale motions have to be modeled.
The models describe the sub-grid-scale stress and the energy transfer between the grid scale and the sub-grid scale.
One of the major models for sub-grid-scale stress is the Smagorinsky-type eddy viscosity \cite{1963_Smagorinsky}.
This model has two parameters: the Smagorinsky constant $C_S$ and the filter function with the filter width.
In the case of homogeneous isotropic turbulence, Lilly \cite{1966_Lilly} determined the Smagorinsky constant  
on the assumption that the grid scale lies within the inertial range of turbulence.
For shear flow turbulence, however, an appropriate value of $C_S$  may change from flow to flow \cite{1970_Deardorff,1998_Hartel}.
In addition, the several types of the LES models have been studied \cite{1991_Germano,2001_Stolz,2005_Kobayashi,2018_Zhuo,2019_Yang_LES}.
We should note that checking the validity of these models, 
they carried out time integration of the NS and LES equations as prior and posterior tests.
These tests have considered whether the LES turbulence gives good approximations for the NS turbulence with respect to 
their statistical properties such as the mean and root-mean-square (RMS) velocity profiles.
Therefore, there is no guarantee that phase space structure of the LES system could imitate that of the NS system.

The present study employs the static Smagorinsky-type eddy viscosity model \cite{1963_Smagorinsky}, 
which is one of the canonical LES systems.
We treat the Smagorinsky constant as a bifurcation parameter and study bifurcation structure of UPOs, 
although the static eddy viscosity is found to cause excessive damping of large-scale fluctuation.
The statistical properties of LES turbulence are known to mimic
those of NS turbulence, and the dynamics of LES flow resembles that of NS flow.
If, in the LES system, we obtain UPOs embedded in the LES turbulent attractor,
these dynamical properties could describe the LES turbulence,
from which we could obtain some perspectives for the NS turbulence.
Having similar motivations, 
Rawat et al. \cite{2015_Rawat} presented an LES steady state arising from NCBW,
which represents the large-scale vortex structure in plane Couette flow.
Hwang et al. \cite{2016_Hwang} studied LES travelling waves in channel flow at very high Reynolds numbers, 
and they suggested that the large-scale structures could be explained using similar mechanisms of the buffer layer. 
Sekimoto \& Jim\'enez \cite{2017_Sekimoto} reported LES steady states in homogeneous shear flow.
The steady states represent vertically localized structure similar to the flow profile of the LES homogeneous shear turbulence.
For LES Taylor-Green flow in the 3D torus, 
van Veen et al. \cite{2019_vanVeen} found a UPO describing pairs of anti-parallel vortices 
which represents the typical coherent structure on the turbulent energy transfer.
Yang et al. \cite{2019_Yang} reported LES travelling waves in channel flow with a narrow spanwise computational domain.
These travelling waves possess a self similarity with respect to the spanwise length at given Reynolds numbers.

\begin{figure}[t]
 \centering
 \includegraphics[width=1.0\textwidth]{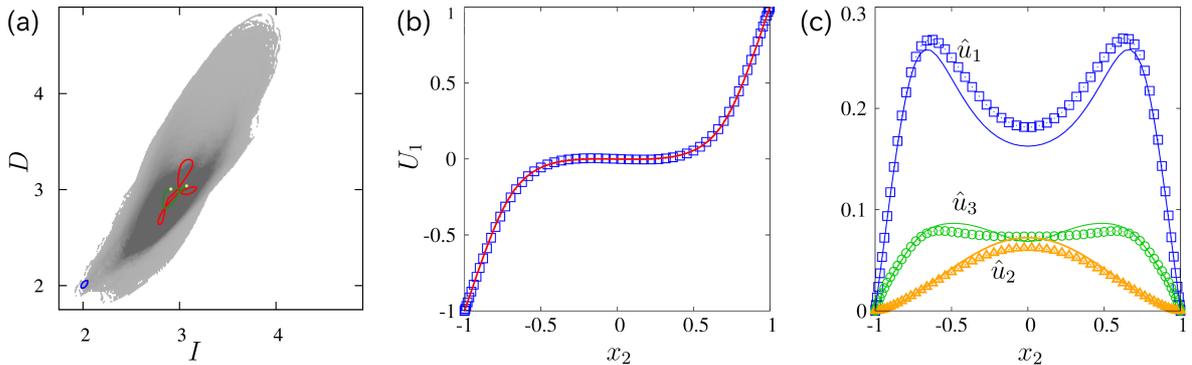}
 \caption{
 Recalculation of Kawahara \& Kida \cite{2001_Kawahara} and newly found UPOs 
 using the number of grid points $(N_1,N_2,N_3)=(48,65,48)$.
 (a) Projection of the UPOs at $Re=400$.
 The horizontal and vertical axes represent the energy injection $I$
 and the energy dissipation $D$, respectively.
 The red and blue lines indicate the solution orbits of the vigorous and gentle UPOs \cite{2001_Kawahara}, respectively.
 Furthermore, the green line means that of the active UPO which is newly examined.
 All the orbits turn clockwise. The gray scale shows the probability density distribution of the NS turbulence.
 The two yellow dots on the green line shows the phases of panels in fig. \ref{fig:profile_re400UG}.
 (b) Mean velocity profile and (c) RMS velocity profiles at $Re=400$:
 the active UPO (lines), the NS turbulence (symbols). 
 }
 \label{fig:projection_mean_RMS_NSRe4e2}
\end{figure}

This paper focuses on two UPOs reported by Kawahara \& Kida \cite{2001_Kawahara} in the NS system.
In present study, we refer to these UPOs as the {\it vigorous}
and {\it gentle} UPOs.
The vigorous UPO is embedded in the turbulent attractor at the Reynolds number $Re=400$.
Here, the Reynolds number is given by the half-distance of the channel height, 
the half-difference of the wall velocities
and the kinematic molecular viscosity.
The statistical properties of the vigorous UPO approximate those of turbulence,
and its dynamics explains the regeneration cycle.
On the other hand, the gentle UPO is an edge state,
which has only one unstable eigenmode.
The edge state separates the turbulent and laminar states in the phase space.
We should note that work by Viswanath \cite{2007_Viswanath} has questioned whether the vigorous UPO really exists at higher spatial resolutions.
Figure \ref{fig:projection_mean_RMS_NSRe4e2} shows our recalculation results.
We have confirmed the existence of the vigorous UPO using the higher resolution than Kawahara \& Kida's \cite{2001_Kawahara},
but in fig. \ref{fig:projection_mean_RMS_NSRe4e2} (a) we could identify the slightly different shape of the projection orbit of the vigorous UPO 
from the original resolution.
On the branch of the gentle UPO we detect a turning point at $Re=236.1$ where the gentle UPO
 connects to an upper branch through the saddle-node bifurcation.
It now turns out that the gentle UPO is a lower branch of the
saddle-node pair, because it is closer to the laminar state than the upper one.
Hereafter, we refer to the UPO on the upper branch of this pair as the {\it active} UPO,
while only for the lower branch of this pair we use the gentle UPO.
We stress that the active UPO is also embedded in the turbulent attractor at $Re=400$.
The mean and RMS velocity profiles of this UPO approximate those of turbulence 
(figs. \ref{fig:projection_mean_RMS_NSRe4e2} (b) and (c)).
Therefore, we expect that not only the vigorous UPO but also the active UPO 
depict representative motions of turbulence at higher Reynolds numbers. 
The present study considers bifurcation structure of 
the vigorous, gentle and active UPOs in the LES system.

The remaining of this paper is organized as follows.
The second section fixes the parameters and explains the numerical methods.
In the third section we present bifurcation structure of the UPOs.
Treating the Smagorinsky constant as a bifurcation parameter,
we discover that the vigorous UPO bifurcates from the active UPO in the LES system.
At the moderately high Reynolds number,
we track the branch of the LES active UPO with changing not only the Smagorinsky constant 
but also the number of grid points and so the filter width.
The bifurcation structure seems similar for different resolutions.
These branches can be rescaled using the Smagorinsky mixing length. 
As the Smagorinsky constant decreases,
the LES active UPO possesses the spanwise vortices created by the streak instability.
The spanwise vortices are stretched by the streamwise vortices, leading to 
the high-Reynolds-stress objects attached to the walls.
These results suggest that the spanwise vortices play a prominent role in
the turbulent momentum transfer as in developed wall turbulence.
The fourth section is devoted to conclusions.

\section{Setup}

We consider the incompressible LES equation with the Smagorinsky
model \cite{1970_Deardorff,1982_Moin}:
\begin{eqnarray*}
 \frac{\partial u_i}{\partial t}+u_j\partial_ju_i&=&-\partial_ip+
  2\partial_j\left\{\left(\frac{1}{Re}+\nu_e\right)S_{ij}\right\},\\
  \partial_iu_i&=&0,
\end{eqnarray*}
where $t$ is the time, $\vec{u}=(u_1,u_2,u_3)$ is the velocity, and $p$ is the kinematic pressure.
Here, $\partial_i=\partial/\partial x_i$ represents the spatial derivative with respect to
$x_i$, where $\vec{x}=(x_1,x_2,x_3)$ indicates the streamwise,
wall-normal, and spanwise coordinates, respectively,
and $S_{ij}=(\partial_i u_j+\partial_j u_i)/2$ is the strain rate tensor.
Quantities have been normalized by the half-distance between the walls $h$
and the half-difference of the wall velocities $U$,
so that the Reynolds number is defined as $Re=Uh/\nu$,
where $\nu$ is the kinematic molecular viscosity.
The dimensionless eddy viscosity $\nu_e$ is defined as
\begin{eqnarray*}
  \nu_e=\{C_S\Delta(x_2)f_S(x_2)\}^2|S|,
\end{eqnarray*}
where $|S|=\sqrt{2S_{ij}S_{ij}}$.
Moreover, $C_S$ denotes the Smagorinsky constant, whose typical value
seems to be in the range of $0.05$ to $0.1$ \cite{1970_Deardorff,1998_Hartel,2015_Rawat}.
The filter width $\Delta(x_2)=\{\Delta_1\Delta_2(x_2)\Delta_3\}^{1/3}$
is estimated in terms of the grid spacing, $\Delta_1$, $\Delta_2(x_2)$ and $\Delta_3$,
in the streamwise, wall-normal, and spanwise directions.
In addition, $f_S(x_2)=1-\exp(-x_2^+/A^+)$ is the van Driest damping function with $A^+=25$ \cite{1956_Driest}.
The $^+$ superscript means quantities rescaled
with the friction velocity $u_\tau$ and the kinematic molecular viscosity.

We impose periodic boundary conditions in the streamwise and spanwise
directions and the non-slip and impermeability conditions at the
walls: 
$u_i(x_1+L_1,x_2,x_3)=u_i(x_1,x_2,x_3+L_3)=u_i(x_1,x_2,x_3);$ 
$u_1(x_1,\pm1,x_3)=\pm1;$ 
$u_2(x_1,\pm1,x_3)=u_3(x_1,\pm1,x_3)=0.$
We set the streamwise and spanwise periods to
$L_1=1.755\pi$ and $L_3=1.2\pi$, which are a minimal flow unit at low Reynolds numbers \cite{1995_Hamilton, 2001_Kawahara}.

The Chebyshev--Fourier--Galerkin spectral method is employed.
We take the number of the grid points as $N_1\geq 3K+1,N_2=L+1,N_3\geq3M+1$,
and increase the resolutions up to $(N_1,N_2,N_3)=(50,65,50)$.
Here, $K$, $L$, and $M$ denote
the streamwise, wall-normal, and spanwise truncation mode numbers, respectively.
In addition, in our numerical setup,
the implicit filtering is employed in the wall-normal direction, 
whereas in the wall-parallel directions the sharp cutoff filter is applied with the filter width equal to the grid size.
Therefore, the wall-normal filter width $\Delta_2(x_2)$ is given by the grid spacing of the Gauss-Lobatto points,
while the streamwise and spanwise ones are equally spacing $\Delta_1=L_1/N_1$ and $\Delta_3=L_3/N_3$.
The time integration is carried out using the second-order Crank--Nicolson and Adams--Bashforth methods.
In order to obtain UPOs, we use the Newton--GMRES method with the stopping condition expressed as
\begin{eqnarray*}
  \frac{||\vec{\phi}_T(\vec{y})-\vec{y}||}{||\vec{y}||}<10^{-5}.
\end{eqnarray*}
Here, $\vec{y}\in\mathbb{R}^{D}(D=4(2K+1)(L-2)(2M+1)+2(L-2))$ is the state vector
given by the real and imaginary parts of the spectral coefficients,
$\vec{\phi}_T(\vec{y})$ is the time-$T$ map defined by
the time integration up to $t=T$ from the initial condition $\vec{y}$,
and $||\vec{y}||=(\vec{y},\vec{y})^{1/2}$ is the Euclidean norm (see \cite{2016_Sasaki}).
The GMRES method is a Krylov subspace method based on sparsity of simultaneous equations.
As the number of grid points increases the dimension of Krylov subspace increases, implying that the UPO computation is numerically tough.
For instance, in order to obtain a UPO at $Re=2000$ and $C_S=0.2598$,
we carry out 20 steps of the Newton iteration with the dimension of the Krylov subspace about up to 80 for each iteration.
Here,  we use the UPO at $Re=2000$ and $C_S=0.2599$ as the initial guess. 
This calculation takes about three days using our computational resource.

\begin{figure}
 \centering
 \includegraphics[width=1.0\textwidth]{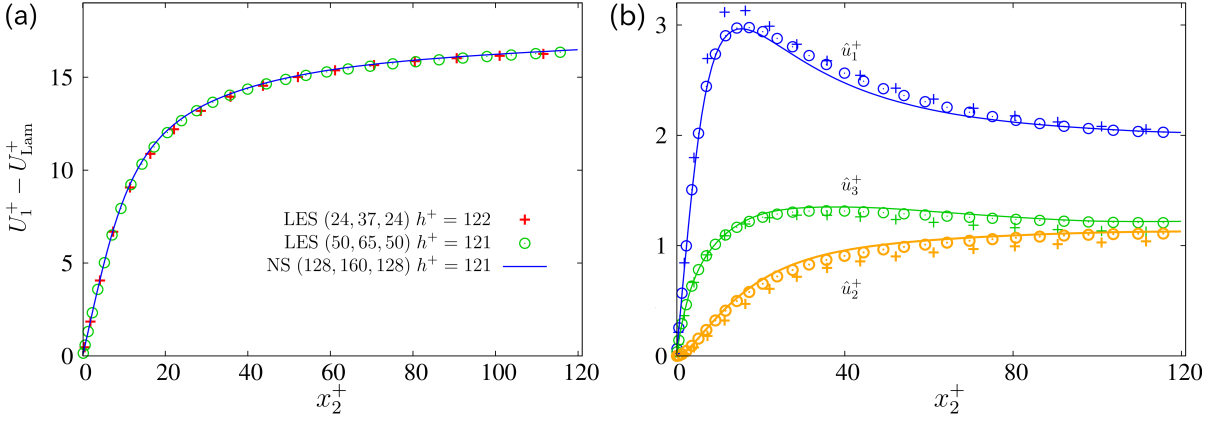}
 \caption{
 Statistical properties of LES and NS turbulence at $Re=2000$ in our numerical computation.
 (a) Mean velocity profile. The linear laminar profile $U_{\rm Lam}^+$ has been subtracted. (b) RMS velocity profiles. 
 The lines represent the NS turbulence for $h^+=121$, while the crosses and open circles denote the LES turbulence at 
 $C_S=0.1$ using the number of grid points $(N_1,N_2,N_3)=(24,37,24), (h^+=122)$ and $(50,65,50), (h^+=121)$, respectively.
 In the panel (b), the blue, orange and green colors indicate
 the RMS velocities $\hat{u}_1^+$, $\hat{u}_2^+$ and $\hat{u}_3^+$, respectively.
 }
 \label{fig:Mean-RMS-NS-LES-Re2e3}
\end{figure}

Before presenting bifurcation structure of LES UPOs, 
we consider an optimal Smagorinsky constant in our numerical setup.
H\"{a}rtel \& Kleiser \cite{1998_Hartel} determined optimal Smagorinsky constants at different Reynolds numbers and spatial resolutions, 
taking into account equilibrium states on the overall exchange of the energy between the grid and sub-grid scales. 
Gullbrand \& Chow \cite{2003_Gullbrand} studied numerical error arising from different discretization methods, 
and they reported that the explicit filtering gives better approximation for the statistical properties.
In this study, however, we only follow work by Deardorff \cite{1970_Deardorff} where he used $C_S=0.1$.
We have carried out time integration of LES and NS equations 
for turbulent flows in the same domain size at several values of the Reynolds numbers.
Figure \ref{fig:Mean-RMS-NS-LES-Re2e3} shows the mean and RMS velocity profiles of LES and NS turbulence at $Re=2000$.
These profiles of the LES turbulence, which are obtained using $C_S=0.1$ and the number of grid points $(N_1,N_2,N_3)=(24,37,24)$,
 resemble those of NS turbulence.
In the case of $Re\leq 1000$, $(N_1,N_2,N_3)=(24,33,24)$ is sufficient 
to achieve the good approximations for these profiles of the NS system.

\section{Bifurcation diagram of LES UPOs}

\begin{figure}[t]
 \centering
 \includegraphics[width=1.0\textwidth]{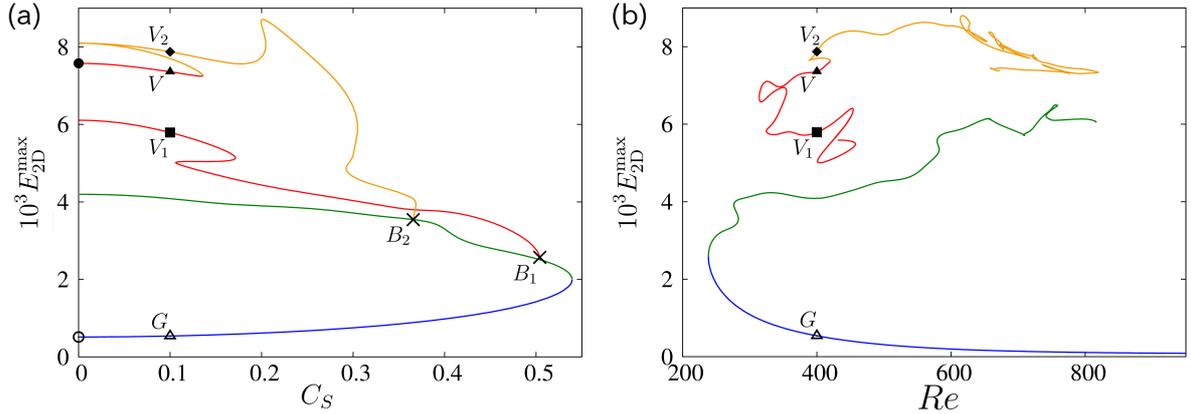}
 \caption{
 Bifurcation diagram arising from the NS vigorous and gentle UPOs: 
 (a) $Re=400$, (b) $C_S=0.1$.
 The solid and open circles represent the vigorous and gentle UPOs
 in the NS system ($C_S=0$) \cite{2001_Kawahara}, respectively.
 The blue and green lines represent the branches of the LES gentle and active UPOs 
 arising through the saddle-node bifurcation. 
 Additionally, the red line indicates the branch stemming from the NS vigorous UPO, while the orange 
 line means the upper branch bifurcating from the branch of the LES vigorous UPO at the turning point.
 The solid triangles, solid squares, solid diamonds, and open triangles indicate the LES UPOs at $Re=400$ and $C_S=0.1$, 
 which we call $V$, $V_1$, $V_2$, and $G$, respectively.
 The crosses on the branch of the LES active UPO in (a) 
 represent the bifurcation points at $C_S=0.5040$ and 0.3657, which we refer to as $B_1$ and $B_2$.
 In order to avoid needless confusion we omit numerous turning points and branches.
 }
 \label{fig:bif_re400cs0.1}
\end{figure}

Let us start by examining LES UPOs with a fixed number of grid points,
$(N_1,N_2,N_3)=(24,33,24)$ and so the filter width. 
We track the branches of the vigorous and gentle UPOs in the LES system and try to increase the Reynolds number. 
Figure \ref{fig:bif_re400cs0.1} describes the bifurcation diagram, where
the vertical axis indicates the maximal value of the cross-flow energy defined as
\begin{eqnarray*}
  E^{\max}_{\mathrm{2D}}=\max_{0\leq t\leq T}\int \frac{u_2^2+u_3^2}{2} \frac{\mathrm{d}V}{2L_1L_3}.
\end{eqnarray*}
The vigorous and gentle UPOs have two symmetries\cite{2001_Kawahara}:
\begin{eqnarray*}
 \Sigma_1&:&(u_1,u_2,u_3)(x_1,x_2,x_3,t)\rightarrow (u_1,u_2,-u_3)(x_1+L_1/2,x_2,-x_3,t),\\
 \Sigma_2&:&(u_1,u_2,u_3)(x_1,x_2,x_3,t)\rightarrow(-u_1,-u_2,u_3)(-x_1,-x_2,x_3+L_3/2,t).
\end{eqnarray*}
In addition, the gentle UPO possesses the following symmetry:
\begin{eqnarray*}
  \Sigma_3:(u_1,u_2,u_3)(x_1,x_2,x_3,t)\rightarrow( u_1, u_2,u_3)(x_1+L_1/2,x_2,x_3,t+T/2).
\end{eqnarray*}
At $Re=400$ as the Smagorinsky constant increases from $C_S=0$ (fig.~\ref{fig:bif_re400cs0.1} (a)),
in the phase space of the LES system, the NS UPOs continuously connect to the LES UPOs. 
Here, we refer to the LES UPOs at $C_S=0.1$ arising from the NS vigorous and gentle UPOs ($C_S=0$) as $V$ and $G$, respectively.
Using the shooting method \cite{2003_Toh_Itano}, we have confirmed that
the LES gentle UPO $G$ is also an edge state as in the NS system.
Tracking the branches originating from $V$ and $G$, we detect two turning points at $C_S=0.1358$ and $0.5393$,
where the corresponding upper branch arises from each branch.
We refer to the solution of the upper branch of the saddle-node pair arising from $G$
as the LES active UPO, while the lower one is the LES gentle UPO.
Note that the branch of the LES active UPO connects to the NS active UPO at $C_S=0$.
On the branch of the LES active UPO, we detect two bifurcation points at $C_S=0.5040$ and 0.3657,
which we call $B_1$ and $B_2$, respectively.
At each bifurcation point, a UPO bifurcates with breaking the symmetry $\Sigma_3$.
Tracking the branch bifurcating from $B_1$, we find an LES UPO at $C_S=0.1$, which we refer to as $V_1$. 
Next, changing the Reynolds number with fixing $C_S=0.1$ (fig.~\ref{fig:bif_re400cs0.1} (b)), 
we figure out that the branch stemming from $V_1$ connects with $V$.
Moreover, on the branch originating from $B_2$ we also identify an LES UPO $V_2$ at $C_S=0.1$ and the connection from $V_2$ to $V$ 
by varying the Reynolds number. 
We determine that in the phase space of the LES system, 
the gentle UPO connects to the vigorous UPO through the branches stemming from $B_1$ and $B_2$. 
Note that Kawahara and Kida \cite{2001_Kawahara} separately examined the vigorous and gentle UPOs in the NS system,
and therefore, the relevance between them was unknown.
This result suggests that introducing the eddy viscosity model has led to
non-trivial solutions and their relationship, and that the eddy viscosity could be a useful homotopy parameter
for searching novel UPOs to the NS equations.

In fig.~\ref{fig:bif_re400cs0.1} (b), as the Reynolds number increases we identify the appearance of
complicated behavior of the upper  branches and resulting several turning points especially at $600\leq Re \leq 800$.
This implies an insufficient resolution.
In addition, we confirmed that the magnitude of the eddy viscosity is less than
approximately 10\% of that of the kinematic molecular viscosity.
This indicates that the eddy viscosity does not work well.
The assumption of the LES models is that the grid scale lies within the inertial range of turbulence.
At low Reynolds numbers, however, the sub-grid scale is comparable to the viscous scale, and
therefore, the assumption is not valid.  
In this sense, the failure of the eddy viscosity model is reasonable. 
Then, we stopped tracking the upper branches, but pursue the  branch
of the LES gentle UPO up to $Re=2000$.
We note that it is easy to track the lower branch without the eddy viscosity, because the lower-branch 
solution does not have small-scale structures. As shown in the introduction, the active UPO, 
the upper branch arising from the gentle UPO, approximates the mean and RMS velocity profiles of NS turbulence at $Re=400$. 
We expect that the active UPO could also depict turbulent motions 
at high Reynolds numbers. Using the Smagorinsky constant as the bifurcation parameter, we 
investigate the branch of the LES gentle UPO at $Re=2000$ and find out the LES active UPO arising from the LES gentle UPO 
through the saddle-node bifurcation.

\begin{figure}[t]
 \centering
 \includegraphics[width=1.0\textwidth]{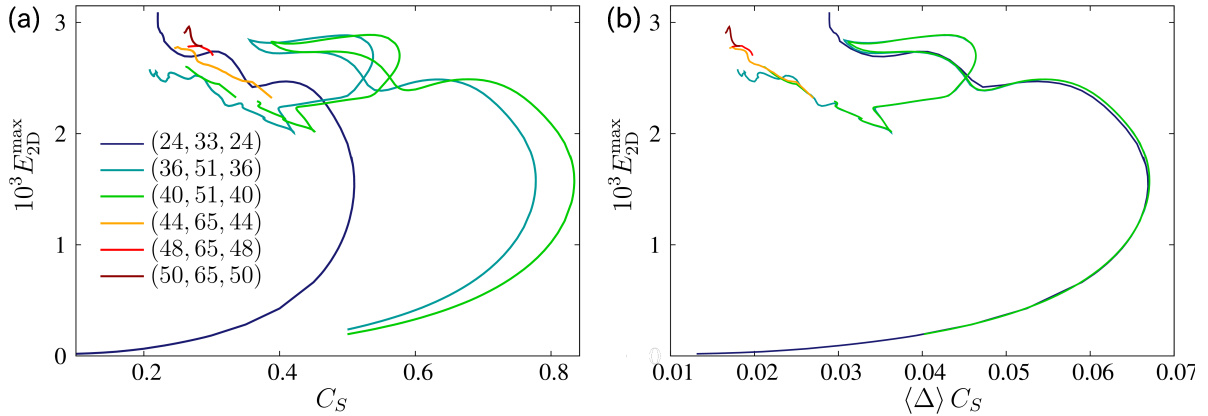}
 \caption{
 Bifurcation diagram of the LES gentle and active UPOs at $Re=2000$.
 The colored lines, from dark-blue to dark-red, represent the sets of the number of grid points
 from $(N_1,N_2,N_3)=(24,33,24)$ to $(50,65,50)$, and so the filter widths.}
 \label{fig:bif_re2000-DeltSmag}
\end{figure}
We track the branches of the LES gentle and active UPOs changing the number of grid points and so the filter width. 
Figure \ref{fig:bif_re2000-DeltSmag} shows the bifurcation structure at $Re=2000$. 
Using the shooting method we confirm that the LES gentle UPO at $Re=2000$ and $C_S=0.1$ is an edge state as in the NS system at $Re=2000$.
Changing the Smagorinsky constant we find that the LES active UPO arises as the saddle-node bifurcation pair with the LES gentle UPO 
at a finite Smagorinsky constant. Moreover, 
the branches exhibit similar behaviors for various resolution (fig.~\ref{fig:bif_re2000-DeltSmag}(a)). 
We discover that these branches can be rescaled using the Smagorinsky mixing length $\braket{\Delta}C_S$.
Here, $\braket{\Delta}$ indicates the averaged filter width:
\begin{eqnarray*}
  \braket{\Delta}=\left(\frac{2L_1L_3}{N_1(N_2-1)N_3}\right)^{1/3}.
\end{eqnarray*}
If we use the sufficient number of grid points, 
the bifurcation structure is robust against the resolution 
and the eddy viscosity does not violate the phase space structure of the LES system.
In order to check the validity of sub-grid-scale models, 
the statistical properties of LES turbulence have been examined.
Using the UPO analysis we could assess proper resolution with respect to the phase space structure in the LES system.
Note that the saddle-node bifurcation point of the pair of the LES gentle and active UPOs is at $\braket{\Delta}C_S\simeq0.067$.
We study the LES active UPO by decreasing the Smagorinsky constant.

\begin{figure}
 \centering
 \includegraphics[width=1.0\textwidth]{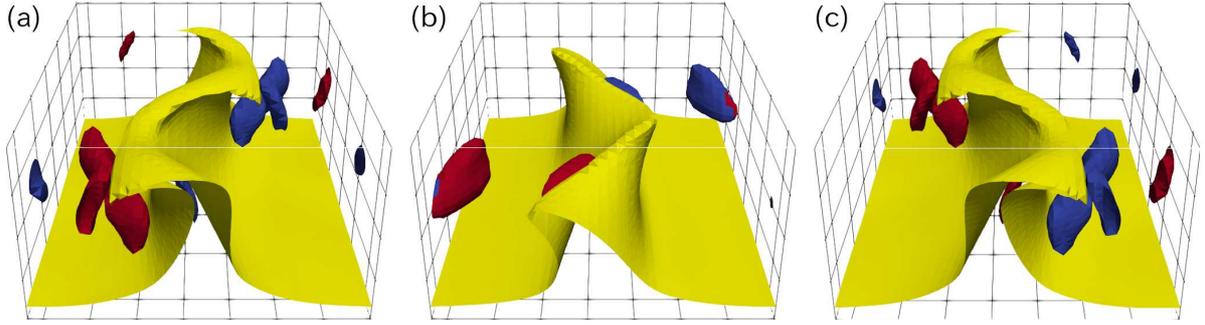}
 \caption{
 Snapshots of the LES active UPO of the period $T=93.72$
 at $Re=2000$ and $\braket{\Delta}C_S=0.05$ ($C_S=0.5797,\;(N_1,N_2,N_3)=(36,51,36),\;h^+=136$): 
 (a) $t=0$, (b) $t=T/4$ and (c) $t=T/2$.
 The yellow object denotes the isosurface of the streamwise velocity $u_1=0.15\;(u^+=12.5)$, while 
 the red and blue objects represent the vortical structure visualized by the isosurfaces of the second 
 invariant of velocity gradient tensor $Q=0.05$ with the positive and negative streamwise vorticity.
 }
 \label{fig:profile_re2000dcs0.05} 
\end{figure}
\begin{figure}
 \centering
 \includegraphics[width=1.0\textwidth]{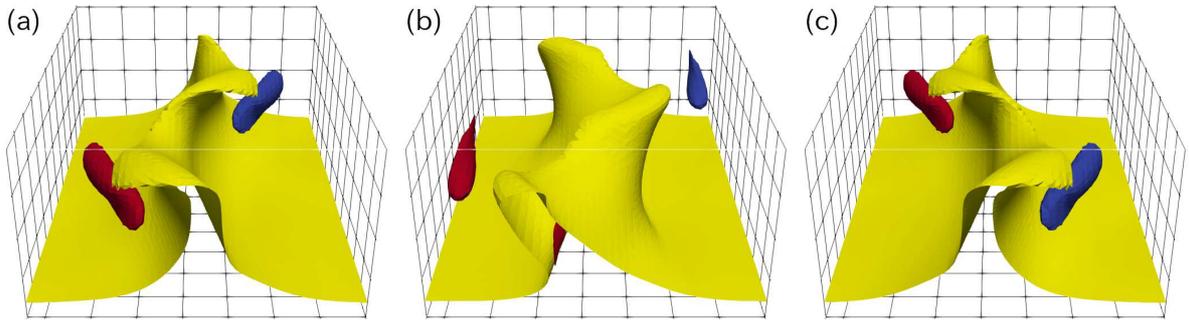}
 \caption{
 Snapshots of the NS active UPO of the period $T=62.13$ at $Re=400\;(h^+=35.0)$: (a) $t=0$, (b) $t=T/5$ and (c) $t=T/2$.
 The definition of isosurfaces is the same as fig.~\ref{fig:profile_re2000dcs0.05}, but $u_1=0.2$ and $Q=0.2$. 
 See also fig.~\ref{fig:projection_mean_RMS_NSRe4e2}.
 }
 \label{fig:profile_re400UG}
\end{figure}
The LES active UPO is embedded in the LES turbulent state at $Re=2000$.
Figure \ref{fig:profile_re2000dcs0.05} shows representative snapshots
of the LES active UPO at $Re=2000,\;\braket{\Delta}C_S=0.05\;(C_S=0.5797,\;(N_1,N_2,N_3)=(36,51,36))$.
The spatial structure exhibits the a pair of positive and negative streamwise vortices and the meandering streak. 
Figure \ref{fig:profile_re400UG} displays the snapshots of the NS active UPO at $Re=400$ for comparison purposes.
The flow structure of the LES active UPO resembles the NS one, although the dissipation mechanism of the LES system
is different from that of the NS system.
We add that the energy injection and the cross-flow energy of the LES active UPO are about twice
larger than those of the NS UPO.
This result implies that the eddy viscosity in the over-damped LES system quenches the small-scale vortices, 
so the flow structures of them seem similar.

\begin{figure}
 \centering
 \includegraphics[width=0.4\textwidth]{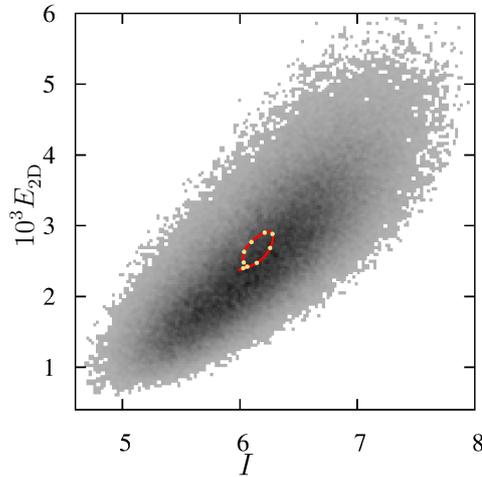}
 \caption{
 Projection of the LES active UPO at $Re=2000$ and $\braket{\Delta}C_S=0.01658$ 
 $(C_S=0.2598,\;(N_1,N_2,N_3)=(50,65,50),\; h^+=105)$.
 The horizontal and vertical axes mean the energy injection $I$ and the cross-flow energy $E_{\rm 2D}$.
 The red line indicates the solution orbit of the LES active UPO.
 The gray scale shows the probability density distribution of the LES turbulence for $h^+=110$.
 The yellow dots represent the phases of panels (a)-(i) in fig. \ref{fig:profile_re2e3cs2598e-4}.
 The orbit turns counter clockwise.
 }
 \label{fig:projection_Re2e3Cs2598e-4}
\end{figure}
\begin{figure}
 \centering
 \includegraphics[width=1.0\textwidth]{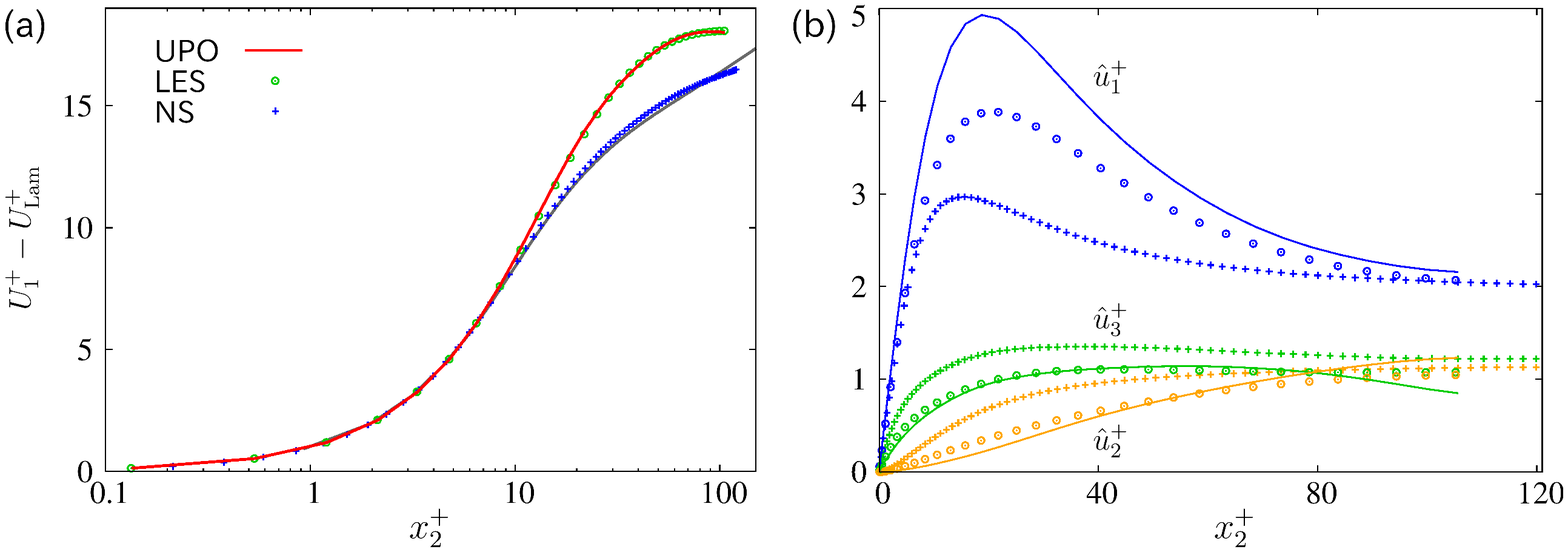}
 \caption{
 (a) Mean velocity and (b) RMS velocity at $Re=2000$. 
 The lines mean the LES active UPO at $Re=2000$ and $\braket{\Delta}C_S=0.01658$
 $(C_S=0.2598,\;(N_1,N_2,N_3)=(50,65,50),\; h^+=105)$, while 
 the open circles represent the LES turbulence $(h^+=110)$ at the same parameters. 
 Additionally, the crosses signify the NS turbulence in our numerical computation, already shown in fig. \ref{fig:Mean-RMS-NS-LES-Re2e3}.
 In the panel (a), the gray line indicates the NS turbulence for $h^+=550$ of the large domain \cite{2014_Avsarkisov},
 whereas in the panel (b), the definition of colors is the same as fig. \ref{fig:Mean-RMS-NS-LES-Re2e3} (b).
 }
 \label{fig:mean-rms}
\end{figure}
As the Smagorinsky constant decreases using the low resolutions, 
we identify the appearance of complicated behavior on the upper branch, again. 
Therefore, we try to decrease the Smagorinsky mixing length and
increase the number of the grid points till the practical numerical cost.
We track the branch down to $\braket{\Delta}C_S=0.01658\;(C_S=0.2598, (N_1,N_2,N_3)=(50,65,50))$.
Figure \ref{fig:projection_Re2e3Cs2598e-4} presents the projection of the solution orbit 
of the LES active UPO on the plane spanned by the energy injection $I$ and the cross-flow energy $E_{\rm 2D}$, where 
$I=\iint (\partial u_1/\partial x_2|_{x_2=1}+\partial u_1/\partial x_2|_{x_2=-1})\mathrm{d}x_1\mathrm{d}x_3/2L_1L_3$.
The solution orbit of the LES active UPO seems to be embedded in the LES turbulent attractor.
Moreover, figure \ref{fig:mean-rms} plots the mean and RMS velocity profiles.
We could find that the mean and RMS velocity profiles of the LES active UPO nicely fit those of the LES turbulence. 
However, these profiles of the LES turbulence do not approximate those of the NS turbulence 
because of the slightly large Smagorinsky constant (see also fig. \ref{fig:Mean-RMS-NS-LES-Re2e3}).
We expect that the LES active UPO provides some insights into the dynamics of NS turbulence despite these discrepancies.

\begin{figure}
 \centering
 \includegraphics[width=1.0\textwidth]{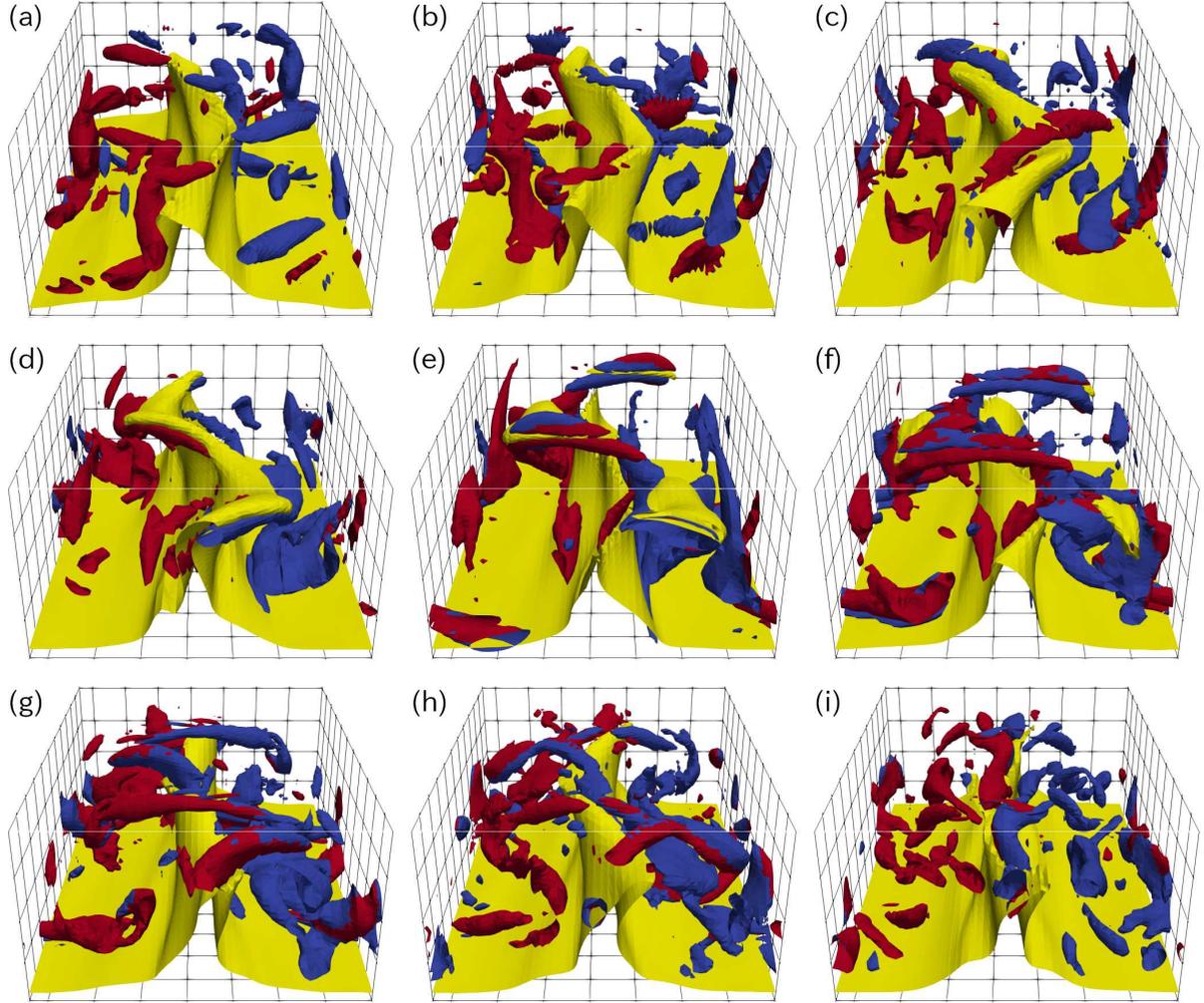}
 \caption{
 Snapshots of the LES active UPO of the period $T=79.25$ 
 at $Re=2000$ and $\braket{\Delta}C_S=0.01658\;(C_S=0.2598, (N_1,N_2,N_3)=(50,65,50),\;h^+=105)$.
 Time elapses from (a) to (i) by $T/20$.
 The definition of isosurfaces is the same as fig.~\ref{fig:profile_re2000dcs0.05}, but $u_1=0.2\;(u^+=21.6)$ and $Q=0.1$.
 }
 \label{fig:profile_re2e3cs2598e-4}
\end{figure}
\begin{figure}
 \centering
 \includegraphics[width=1\textwidth]{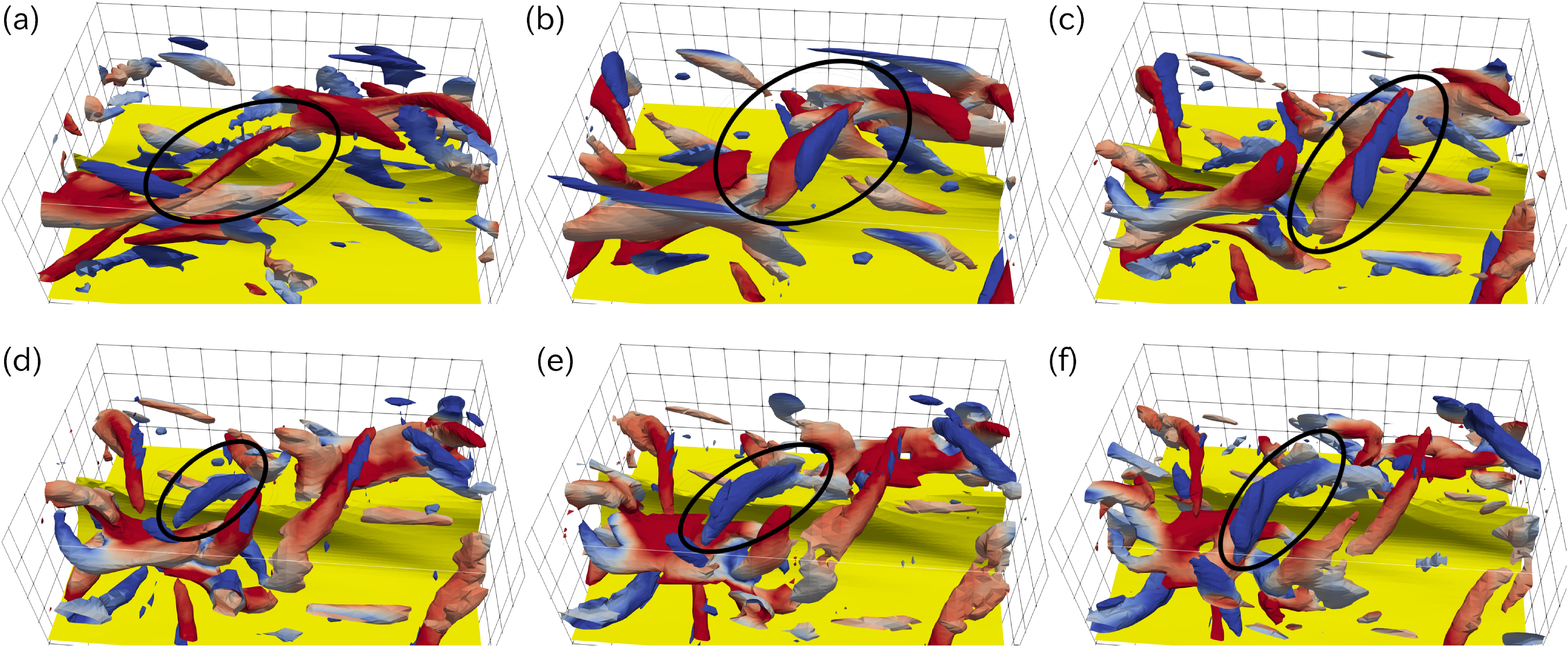}
 \caption{
 Development of spanwise vortices.
 The yellow objects are the isosurfaces of the streamwise velocity $u_1=0.3$,
 while the blue to red objects denote $Q=0.1$ within the range of $-0.75\leq\omega_3\leq0.75$.
 Time elapses from (a) to (c) by $T/20\;(3T/20\leq t\leq5T/20)$
 and from (d) to (f) by $T/40\;(55T/200\leq t\leq65T/200)$.
 The black circles are used for emphasis on the spanwise vortices.}
 \label{fig:profile_re2e3cs2598e-4-zoom}
\end{figure}

Figure \ref{fig:profile_re2e3cs2598e-4} shows the typical snapshots of the LES active UPO. 
We could find more prominent streak motions and vortical structures than in fig. \ref{fig:profile_re2000dcs0.05}. 
From fig. \ref{fig:profile_re2e3cs2598e-4} (a) to (d), 
the streak, the yellow objects, becomes wavy driven by the streamwise vortices, and 
from fig. \ref{fig:profile_re2e3cs2598e-4} (e) to (g), 
we can identify the spanwise vortices in the central region of the channel. 
The wavy streak and spanwise vortices seem to be caused by the streak instability \cite{2003_Kawahara}.  
Figure \ref{fig:profile_re2e3cs2598e-4-zoom} represents the detail of development of the spanwise vortices. 
The pair of the streamwise vortices stretches and intensifies the spanwise vortices. 
We should stress that to our knowledge a spanwise vortex has not been reported through the UPO analysis of wall flows. 
From fig \ref{fig:profile_re2e3cs2598e-4} (h) to (i), 
it can be seen that the spanwise vortices reach the vicinity of the other walls and eventually decay.
\begin{figure}
 \centering
 \includegraphics[width=1\textwidth]{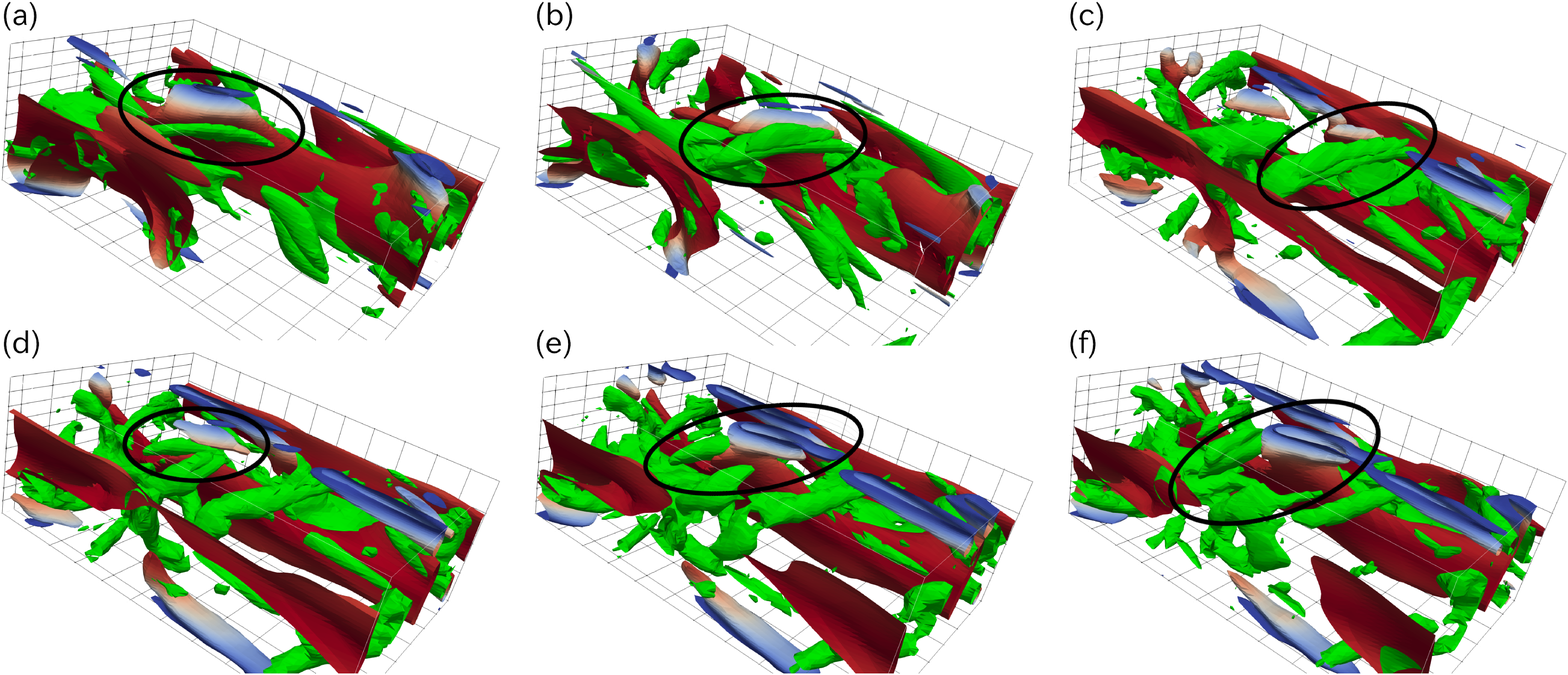}
 \caption{
 Reynolds shear stress $-u'_1u'_2$ of the LES active UPO.
 The blue to red objects indicate the isosurface
 $|u'_1u'_2|/(\hat{u}_1\hat{u}_2)=1.75$ 
 where the color means the value of the Reynolds shear stress $-u'_1u'_2$ from 0 to $0.014$.
 The green surfaces indicate $Q=0.1$.
 The time series of (a) to (c) and of (d) to (f) are the same as fig.\ref{fig:profile_re2e3cs2598e-4-zoom}
 }
 \label{fig:profile_tanRe}
\end{figure}

The streak dynamics represents the streamwise-momentum transfer through turbulent fluctuation.
In order to figure out the relation of the turbulent momentum transfer and the spanwise vortical structure, 
we consider the momentum transport by turbulent fluctuations.
In the averaged equation of the streamwise velocity,
the derivative of the Reynolds shear stress $-u_1'u_2'$
is the source term for the turbulent fluctuation.
Here, $u_i'$ is the fluctuation velocity.
Hence the large Reynolds stress means the prominent modulation
of the turbulence from the mean flow.
Figure \ref{fig:profile_tanRe} shows the isosurfaces of
the rescaled Reynolds shear stress \cite{2012_Lozano}
\begin{eqnarray*}
 \frac{|u_1'(\vec{x})u_2'(\vec{x})|}{\hat{u}_1(x_2)\hat{u}_2(x_2)}=1.75,
\end{eqnarray*}
where $\hat{u}_i$ signifies the RMS velocity.
The high-Reynolds-stress objects are embedded in the streaks and
sandwiched between the streamwise vortices.
When the spanwise vortices are created in the central region of the channel, 
the high-Reynolds-stress objects attach to the walls. 
These results imply that the appearance of the spanwise vortices is the trigger of the significant momentum transfer.
In fully developed turbulence, the spanwise vortices 
have been observed mainly in the overlap region apart from the wall \cite{2006_Wu}.
We expect that the spanwise vortices of the LES active UPO could represent the key events of
developed turbulence, although it is not clear whether the dynamics of the present UPO is strictly 
consistent with that in near-wall turbulence at higher Reynolds numbers.

\section{Conclusions}

Toward perspectives of dynamical properties of turbulence,
we herein document the bifurcation structure of unstable periodic orbits (UPOs).
In order to decrease the degrees of freedom
we consider the large-eddy simulation (LES) with the Smagorinsky-type eddy viscosity model.
Treating the Smagorinsky constant as a bifurcation parameter, 
we study bifurcation structure of UPOs at moderately high Reynolds number.

We demonstrate that the vigorous UPO bifurcates from the branch of 
the newly examined active UPO arising from the gentle UPO as the saddle-node pair in the LES system.
We note that the vigorous and gentle UPOs were separately unveiled 
by extraction from the chaotic attractor \cite{2001_Kawahara}.
This result suggests that 
the eddy viscosity model could present a useful homotopy parameter for searching UPOs, 
and we could seek novel nonlinear solutions and their relationship.

As the Reynolds number slightly increases, the upper branches have numerous turning points. 
In the LES models, we assume that the grid scale lies within the inertial range of turbulence. 
At low Reynolds numbers, however, the sub-grid scale is comparable to the viscous scale, 
and the model assumption is not valid.
In this sense, the failure of the eddy viscosity model is reasonable.
We stopped tracking the upper branches, but continue to track the lower branch, i.e., the LES gentle UPO.

It is easy to track the branch of the gentle UPO without eddy viscosity, 
because the lower-branch solution does not have small-scale structures.
Since, as shown in the introduction, the active UPO of the saddle-node pair with the gentle UPO describes the low Reynolds turbulence, 
we expect that the LES active UPO could represent the high Reynolds turbulence.
Using the Smagorinsky constant as the bifurcation parameter at the moderately high Reynolds number, 
we find out the LES active UPO arising from the LES gentle UPO through the saddle-node bifurcation.
The bifurcation structure of the LES gentle and active UPOs seems similar for different resolutions.
These branches could be rescaled with the Smagorinsky mixing length $\braket{\Delta}C_S$. 
Here $\braket{\Delta}$ is the averaged filter width.
This result suggests that if we use the sufficient number of grid points, 
the eddy viscosity does not violate bifurcation structure.
Using the UPO analysis we could examine adequate resolutions in terms of the phase space structure in the LES system.

At the large Smagorinsky constant, the eddy viscosity quenches the small-scale structure,
so the flow structures of the LES active UPO seems similar to that of the NS active UPO at the low Reynolds number.
When we decrease the Smagorinsky constant, however, we discover that the LES active UPO possesses the spanwise vortices 
which appear in the central region of the channel and seem to be caused by the streak instability \cite{2003_Kawahara}.
The spanwise vortices are stretched by the streamwise vortices, leading to 
the high-Reynolds-stress objects attached to the walls.
This result suggests that the spanwise vortices of the LES active UPO 
enhance the transfer of the streamwise turbulent momentum as in developed wall turbulence.

We should note that the mean and RMS velocities of the LES active UPO do not approximate those of the NS turbulence.
As the Smagorinsky constant decreases we identify a lots of turning points because of the lack of resolution.
We need to increase the number of grid points, but it is practically hard using our present computational resources. 
It may be a future research to develop more powerful numerical methods of the UPO analysis.

At the high Reynolds number, the LES active UPO describes the creation of the spanwise vortices which enhance turbulent momentum transfer.
In fully developed turbulence, the spanwise vortices have been observed mainly in the overlap region apart from the wall \cite{2006_Wu}.
We expect that the spanwise vortices of the LES active UPO could capture the key events of
developed turbulence, although it is not clear whether the dynamics of the present UPO is strictly
consistent with that in near-wall turbulence at higher Reynolds numbers.
We need to further increase the Reynolds number,
and moreover it is required to investigate UPOs for channel flow and turbulent boundary layer.
The spanwise vortices in near-wall turbulence play important roles in the friction drag and heat transfer
\cite{2011_GMayoral,2013_Yamamoto}.
We believe that the dynamics of the spanwise vortices demonstrated by the present UPO 
gives some insights to improve the turbulent control strategies.\\

 This work was partially supported by JSPS KAKENHI Grant Number 50214672.
 The third author is supported by ERC-2014.AdG-669505 Coturb.
\bibliographystyle{jplain}
\bibliography{2021_LESCouette_rev.v1}

\providecommand{\noopsort}[1]{}\providecommand{\singleletter}[1]{#1}
\begin{thebibliography}{10}

\bibitem{2013_Avila}
M.~Avila, F.~Mellibovsky, N.~Roland, and B.~Hof.
\newblock Streamwise-localized solutions at the onset of turbulence in pipe
  flow.
\newblock {\em Phys. Rev. Lett.}, Vol. 110, p. 224502, May 2013.

\bibitem{2014_Avsarkisov}
V.~Avsarkisov, S.~Hoyas, M.~Oberlack, and J.^^c2^^a0P. Garc^^c3^^ada-Galache.
\newblock Turbulent plane {Couette} flow at moderately high {Reynolds} number.
\newblock {\em J. Fluid Mech.}, Vol. 751, p.~R1, 2014.

\bibitem{2014_Brand}
E.~Brand and J.^^c2^^a0F. Gibson.
\newblock A doubly localized equilibrium solution of plane {Couette} flow.
\newblock {\em J. Fluid Mech.}, Vol. 750, p.~R3, 2014.

\bibitem{1997_Clever}
R.~M. Clever and F.~H. Busse.
\newblock Tertiary and quaternary solutions for plane {Couette} flow.
\newblock {\em J. Fluid Mech.}, Vol. 344, pp. 137--153, 1997.

\bibitem{1995_Dauchot}
O.~Dauchot and F.~Daviaud.
\newblock Finite amplitude perturbation and spots growth mechanism in plane
  {Couette} flow.
\newblock {\em Phys. Fluids}, Vol.~7, No.~2, pp. 335--343, 1995.

\bibitem{1970_Deardorff}
J.~W. Deardorff.
\newblock A numerical study of three-dimensional turbulent channel flow at
  large {Reynolds} numbers.
\newblock {\em J. Fluid Mech.}, Vol.~41, No.~2, pp. 453--480, 1970.

\bibitem{2010_Duguet}
Y.~Duguet, P.~Schlatter, and D.~S. Henningson.
\newblock Formation of turbulent patterns near the onset of transition in plane
  {Couette} flow.
\newblock {\em J. Fluid Mech.}, Vol. 650, pp. 119--129, 2010.

\bibitem{2018_Eckhardt}
B.~Eckhardt and S.~Zammert.
\newblock Small scale exact coherent structures at large {Reynolds} numbers in
  plane {Couette} flow.
\newblock {\em Nonlinearity}, Vol.~31, No.~2, pp. R66--R77, jan 2018.

\bibitem{2011_GMayoral}
R.~Garc^^c3^^ada-Mayoral and J.~Jim^^c3^^a9nez.
\newblock Drag reduction by riblets.
\newblock {\em Phil. Trans. R. Soc. A}, Vol. 369, No. 1940, pp. 1412--1427,
  2011.

\bibitem{1991_Germano}
M.~Germano, U.~Piomelli, P.~Moin, and W.~H. Cabot.
\newblock A dynamic subgrid-scale eddy viscosity model.
\newblock {\em Phys. Fluid A}, Vol.~3, pp. 1760--1765, 1991.

\bibitem{2003_Gullbrand}
J.~Gullbrand and F.~K. Chow.
\newblock The effect of numerical errors and turbulence models in large-eddy
  simulations of channel flow, with and without explicit filtering.
\newblock {\em J. Fluid Mech.}, Vol. 495, pp. 323--341, 2003.

\bibitem{1995_Hamilton}
J.~M. Hamilton, J.~Kim, and F.~Waleffe.
\newblock Regeneration mechanisms of near-wall turbulence structures.
\newblock {\em J. Fluid Mech.}, Vol. 287, pp. 317--348, 1995.

\bibitem{1998_Hartel}
C.~H\"{a}rtel and L.~Kleiser.
\newblock Analysis and modelling of subgrid-scale motions in near-wall
  turbulence.
\newblock {\em J. Fluid Mech.}, Vol. 356, pp. 327--352, 1998.

\bibitem{2013_Hirsch}
M.~W. Hirsch, S.~Smale, and R.~L. Devaney.
\newblock {\em Differential Equations, Dynamical Systems, and an Introduction
  to Chaos}.
\newblock Academic Press, 3 edition, 2012.

\bibitem{2016_Hwang}
Y.~Hwang, A.~P. Willis, and C.~Cossu.
\newblock Invariant solutions of minimal large-scale structures in turbulent
  channel flow for {$Re_\tau$} up to 1000.
\newblock {\em J. Fluid Mech.}, Vol. 802, p.~R1, 2016.

\bibitem{1991_Jimenez}
J.~Jim\'{e}nez and P.~Moin.
\newblock The minimal flow unit in near-wall turbulence.
\newblock {\em J. Fluid Mech.}, Vol. 225, pp. 213--240, 1991.

\bibitem{2003_Kato}
S.~Kato and M.~Yamada.
\newblock Unstable periodic solutions embedded in a shell model turbulence.
\newblock {\em Phys. Rev. E}, Vol.~68, p. 025302(R), Aug 2003.

\bibitem{2005_Kawahara}
G.~Kawahara.
\newblock Laminarization of minimal plane {Couette} flow: Going beyond the
  basin of attraction of turbulence.
\newblock {\em Phys. Fluids}, Vol.~17, No.~4, p. 041702, 2005.

\bibitem{2003_Kawahara}
G.~Kawahara, J.~Jim\'{e}nez, M.~Uhlmann, and A.~Pinelli.
\newblock Linear instability of a corrugated vortex sheet ^^e2^^80^^93 a model
  for streak instability.
\newblock {\em J. Fluid Mech.}, Vol. 483, pp. 315--342, 2003.

\bibitem{2001_Kawahara}
G.~Kawahara and S.~Kida.
\newblock Periodic motion embedded in plane {Couette} turbulence: regeneration
  cycle and burst.
\newblock {\em J. Fluid Mech.}, Vol. 449, pp. 291--300, 2001.

\bibitem{2012_Kawahara}
G.~Kawahara, M.~Uhlmann, and L.~van Veen.
\newblock The significance of simple invariant solutions in turbulent flows.
\newblock {\em Annu. Rev. Fluid Mech.}, Vol.~44, pp. 203--225, 2012.

\bibitem{2005_Kawasaki}
M.~Kawasaki and S.I. Sasa.
\newblock Statistics of unstable periodic orbits of a chaotic dynamical system
  with a large number of degrees of freedom.
\newblock {\em Phys. Rev. E}, Vol.~72, p. 037202, Sep 2005.

\bibitem{2005_Kobayashi}
H.~Kobayashi.
\newblock The subgrid-scale models based on coherent structures for rotating
  homogeneous turbulence and turbulent channel flow.
\newblock {\em Phys. Fluids}, Vol.~17, No.~4, p. 045104, 2005.

\bibitem{1966_Lilly}
D.~K. Lilly.
\newblock On the application of the eddy viscosity concept in the inertial
  sub-range of turbulence.
\newblock {\em NCAR Manuscript}, Vol. 123, , 1966.

\bibitem{2020_Linkmann}
M.~Linkmann, F.~Knierim, S.~Zammert, and B.~Eckhardt.
\newblock Linear feedback control of invariant solutions in channel flow.
\newblock {\em J. Fluid Mech.}, Vol. 900, p. A10, 2020.

\bibitem{2012_Lozano}
A.~Lozano-Dur\'{a}n, O.~Flores, and J.~Jim\'{e}nez.
\newblock The three-dimensional structure of momentum transfer in turbulent
  channels.
\newblock {\em J. Fluid Mech.}, Vol. 694, pp. 100--130, 2012.

\bibitem{2019_Lustro}
J.~Lustro, G.~Kawahara, L.~van Veen, M.~Shimizu, and H.~Kokubu.
\newblock The onset of transient turbulence in minimal plane {Couette} flow.
\newblock {\em J. Fluid Mech.}, Vol. 862, p.~R2, 2019.

\bibitem{1982_Moin}
P.~Moin and J.~Kim.
\newblock Numerical investigation of turbulent channel flow.
\newblock {\em J. Fluid Mech.}, Vol. 118, pp. 341--377, 1982.

\bibitem{1990_Nagata}
M.~Nagata.
\newblock Three-dimensional finite-amplitude solutions in plane {Couette} flow:
  bifurcation from infinity.
\newblock {\em J. Fluid Mech.}, Vol. 217, pp. 519--527, 1990.

\bibitem{2015_Okino}
S.~Okino.
\newblock {\em K\^{o}ky\^{u}roku, RIMS, Kyoto University (in Japanese)}, Vol.
  1994, pp. 24--29, 2015.

\bibitem{2015_Rawat}
S.~Rawat, C.~Cossu, Y.~Hwang, and F.~Rincon.
\newblock On the self-sustained nature of large-scale motions in turbulent
  {Couette} flow.
\newblock {\em J. Fluid Mech.}, Vol. 782, pp. 515--540, 2015.

\bibitem{2016_Sasaki}
E.~Sasaki, G.~Kawahara, A.~Sekimoto, and J.~Jim{\'{e}}nez.
\newblock Unstable periodic orbits in plane {Couette} flow with the smagorinsky
  model.
\newblock {\em J. Phys. Conf. Seri.}, Vol. 708, p. 012003, apr 2016.

\bibitem{2017_Sekimoto}
A.~Sekimoto and J.~Jim^^c3^^a9nez.
\newblock Vertically localised equilibrium solutions in large-eddy simulations
  of homogeneous shear^^c2^^a0flow.
\newblock {\em J. Fluid Mech.}, Vol. 827, pp. 225--249, 2017.

\bibitem{1963_Smagorinsky}
J.~Smagorinsky.
\newblock General circulation experiments with the primitive equations: I. the
  basic experiment.
\newblock {\em Mon. Wea. Rev.}, Vol.~91, No.~3, pp. 99--164, 03 1963.

\bibitem{2001_Stolz}
S.~Stolz, N.~A. Adams, and L.~Kleiser.
\newblock An approximate deconvolution model for large-eddy simulation with
  application to incompressible wall-bounded flows.
\newblock {\em Phys. Fluids}, Vol.~13, No.~4, pp. 997--1015, 2001.

\bibitem{2003_Toh_Itano}
S.~Toh and T.~Itano.
\newblock A periodic-like solution in channel flow.
\newblock {\em J. Fluid Mech.}, Vol. 481, pp. 67--76, 2003.

\bibitem{1956_Driest}
E.~R. van Driest.
\newblock On turbulent flow near a wall.
\newblock {\em J. Aeronaut. Sci.}, Vol.~23, No.~11, pp. 1007--1011, 1956.

\bibitem{2019_vanVeen}
L.~van Veen, A.~Vela-Mart\'{\i}n, and G.~Kawahara.
\newblock Time-periodic inertial range dynamics.
\newblock {\em Phys. Rev. Lett.}, Vol. 123, p. 134502, Sep 2019.

\bibitem{2007_Viswanath}
D.~Viswanath.
\newblock Recurrent motions within plane {Couette} turbulence.
\newblock {\em J. Fluid Mech.}, Vol. 580, pp. 339--358, 2007.

\bibitem{1998_Waleffe}
F.~Waleffe.
\newblock Three-dimensional coherent states in plane shear flows.
\newblock {\em Phys. Rev. Lett.}, Vol.~81, p. 4140, Nov 1998.

\bibitem{2018_Zhuo}
Z.~Wang, K.~Luo, D.~Li, J.~Tan, and J.~Fan.
\newblock Investigations of data-driven closure for subgrid-scale stress in
  large-eddy simulation.
\newblock {\em Phys. Fluids}, Vol.~30, No.~12, p. 125101, 2018.

\bibitem{2006_Wu}
Y.~Wu and K.~T. Christensen.
\newblock Population trends of spanwise vortices in wall turbulence.
\newblock {\em J. Fluid Mech.}, Vol. 568, pp. 55--76, 2006.

\bibitem{2013_Yamamoto}
A.~Yamamoto, Y.~Hasegawa, and N.~Kasagi.
\newblock Optimal control of dissimilar heat and momentum transfer in a fully
  developed turbulent channel flow.
\newblock {\em J. Fluid Mech.}, Vol. 733, pp. 189--220, 2013.

\bibitem{2019_Yang}
Q.~Yang, A.~P. Willis, and Y.~Hwang.
\newblock Exact coherent states of attached eddies in channel flow.
\newblock {\em J. Fluid Mech.}, Vol. 862, pp. 1029--1059, 2019.

\bibitem{2019_Yang_LES}
X.~I.~A. Yang, S.~Zafar, J.-X. Wang, and H.~Xiao.
\newblock Predictive large-eddy-simulation wall modeling via physics-informed
  neural networks.
\newblock {\em Phys. Rev. Fluids}, Vol.~4, p. 034602, Mar 2019.

\bibitem{2014_Zammert}
S.~Zammert and B.~Eckhardt.
\newblock Streamwise and doubly-localised periodic orbits in plane {Poiseuille}
  flow.
\newblock {\em J. Fluid Mech.}, Vol. 761, pp. 348--359, 2014.

\end{thebibliography}
\end{document}